\documentclass[twolecolumn]{elsart3p}
\usepackage{amssymb}
\usepackage[dvips]{graphicx}
\usepackage{bm}% bold math
\usepackage{amsmath}
\journal{Physics Procedia}

\begin{document}

\begin{frontmatter}


%5 \author{Name\corref{cor1}\fnref{label2}}

\title{Effect of anisotropic Fermi surface on the flux-flow resistivity under rotating magnetic field}


\author[AA,CC]{Y. Higashi\corauthref{cor1}}
\ead{higashiyoichi@ms.osakafu-u.ac.jp}
%\ead[url]{http://www.nanosq.21c.osakafu-u.ac.jp/ttsl_lab/n_hayashi/Lab_HP/members/higashi/higashi.html}
\author[BB]{Y. Nagai}
\author[BB]{M. Machida}
\author[CC]{N. Hayashi}

\address[AA]{
Department of Mathematical Sciences, Osaka Prefecture University, 1-1 Gakuen-cho, Naka-ku, Sakai 599-8531, Japan
}

\address[BB]{
CCSE, Japan Atomic Energy Agency, 5-1-5 Kashiwanoha, Kashiwa, Chiba 277-8587, Japan
}

\address[CC]{
NanoSquare Research Center (N2RC), Osaka Prefecture University, 1-2 Gakuen-cho, Naka-ku, Sakai 599-8570, Japan
}
\corauth[cor1]{Corresponding author.
N2RC, Osaka Prefecture University, C10 Bldg., 1-2 Gakuen-cho, Naka-ku, Sakai 599-8570, Japan
Tel.: +81-72-254-9829 ; fax: +81-72-254-8203.}

% #######################################################################################
% #######################################################################################
\begin{abstract}
We numerically investigate the effect of in-plane anisotropic Fermi surface (FS)
on the flux-flow resistivity $\rho_{\rm f}$ under rotating magnetic field
on the basis of the quasiclassical Green's function method.
We demonstrate that one can detect the phase in pairing potential of Cooper pair
through the field-angular dependence of $\rho_{\rm f}$
even if the FS has in-plane anisotropy.
In addition,
we point out one can detect the gap-node directions
irrespective of the FS anisotropy
by measuring $\rho_{\rm f}$ under rotating field.
\end{abstract}
% #########################################################################
% #########################################################################
% #########################################################################
% #########################################################################
\end{frontmatter}

% \linenumbers


\section{Introduction}
%\label{}
It is important task to clarify the internal degree of freedom for the orbital part of Cooper pair wave function,
which is the fundamental nature of superconductivity.
We have ever proposed a new experimental method to detect both the phase and the anisotropy of pairing potential.
That is,
we have investigated the in-plane magnetic field-angle dependence of the quasiparticle scattering rate inside a vortex core \cite{higashi2011}
and the flux-flow resistivity $\rho_{\rm f}$ \cite{higashi}.
Through a series of our researches, 
we obtained the knowledge that the field-angle dependence of $\rho_{\rm f}$ is sensitive to the phase of pairing potential
both in the cases of an isotropic and an uniaxially anisotropic Fermi surface (FS).
The field-angle dependence of $\rho_{\rm f}$ has not been investigated yet
for an in-plane anisotropic FS.

In this paper,
we investigate effects of in-plane FS anisotropy on
the field-angle dependence of flux-flow resistivity $\rho_{\rm f}(\alpha_{\rm M})$.
Most materials have the anisotropy in their FS reflecting the anisotropy of crystal structures.
Thus, it is more realistic to investigate $\rho_{\rm f}(\alpha_{\rm M})$ in the case of anisotropic FS.
We consider two model FSs with in-plane anisotropy
and numerically calculate $\rho_{\rm f}(\alpha_{\rm M})$ for those FSs
with changing the anisotropy of FS.
Our numerical results show that one can detect the gap-node direction
by measuring $\rho_{\rm f}$ under rotating magnetic field
even if the FS has an anisotropy.

%So, we also investigate the effect of the large density of states (DOS) on FS on $\rho_{\rm f}(\alpha_{\rm M})$.
%The DOS is inversely proportional to the gradient of the band structure.

% #########################################################################
% #########################################################################
\section{Formulation}
We consider a single vortex at low magnetic field and at low temperature.
The energy dissipation due to the vortex flow comes from non-magnetic impurity scattering within a vortex core.
Two contributions to the flux-flow resistivity $\rho_{\rm f}$ are considered \cite{higashi}.
One is the quasiparticle scattering due to randomely distributed impurities \cite{higashi2011,nagai2010}
and the othter is the energy scale of the quasiparticle (QP) bound states inside a vortex core $\omega_0({\bm k}_{\rm F})$ \cite{higashi,kopnin}.
Note that this energy scale depends on the wave vector \cite{volovik}
within the framework of the quasiclassical theory of superconductivity.

We assume an isotropic vortex characterized by the pair potential given as
$\Delta({\bm r},{\bm k}_{\rm F})=|\Delta({\bm r})| d({\bm k}_{\rm F}) e^{i\phi(|{\bm r}|)}$.
We set $|\Delta({\bm r})|=\Delta_\infty \tanh(|{\bm r}|/\xi)$ as the spatial variation of the pair potential amplitude.
$\Delta_\infty$ is the bulk amplitude,
$\xi$ is the coherence length,
and $\phi(|{\bm r}|)$ is the azimuthal angle in the real space.
$d({\bm k}_{\rm F})$ denotes the anisotropy of the pair potential in the ${\bm k}$-space.
${\bm k}_{\rm F}$ is the Fermi wave vector.

The expression for the flux-flow resistivity $\rho_{\rm f}$ is  given as \cite{higashi}
%%%%%%%%%%%%%%%%%%%%%%%%%%%%%
\begin{equation}
\rho_{\rm f}(T)
\propto
\frac{1}{\left\langle \frac{\displaystyle \omega_0({\bm k}_{\rm F})}{\displaystyle \varGamma(\varepsilon=k_{\rm B}T,{\bm k}_{\rm F})} \right\rangle_{\rm FS}},
\end{equation}
%%%%%%%%%%%%%%%%%%%%%%%%%%%%%
where the integral on a FS with respect to ${\bm k}_{\rm F}$ is
$\left\langle \cdots \right\rangle=(1/{\nu_0})\int dS_{\rm F}/|{\bm v}_{\rm F}({\bm k}_{\rm F})| \cdots$.
The area element on an anisotropic FS is $dS_{\rm F}=|{\bm k}_{\rm F}(\phi_k,\theta_k)|^2 \sin \theta_k d\phi_k d\theta_k$.
The total density of state on a FS is $\nu_0=\int dS_{\rm F}/|{\bm v}_{\rm F}({\bm k}_{\rm F})|$.
The Fermi velocity is ${\bm v}_{\rm F}({\bm k}_{\rm F})={\bm \nabla}_{\bm k} \epsilon({\bm k})|_{{\bm k}={\bm k}_{\rm F}}$.
In this paper, we use a unit system in which $\hbar=1$.
Here,
we assume that the system is in moderately clean regime and that quasiparticles with energy $\varepsilon =k_{\rm B}T$
predominantly contribute to the flux-flow resistivity at the temperature $T$ \cite{kato2000}.
Within the quasiclassical theory,
the momentum dependent interlevel spacing of the vortex bound states $\omega_0({\bm k}_{\rm F})$
is obtained by using Kramer-Pesch approximation \cite{nagai2008} as
%%%%%%%%%%%%%%%%%%%%%%%%%%%%%
$
\omega_0({\bm k}_{\rm F})
=
2 |d({\bm k}_{\rm F})|^2 \Delta^2_\infty
/(|{\bm k}_{\rm F \perp}| |{\bm v}_{\rm F \perp}({\bm k}_{\rm F})|)
$
\cite{higashi,nagai2010}.
%%%%%%%%%%%%%%%%%%%%%%%%%%%%%
Here,
${\bm k}_{\rm F \perp}$ and ${\bm v}_{\rm F \perp}$
are the components of ${\bm k}_{\rm F}$ and ${\bm v}_{\rm F}$
projected onto the plane perpendicular to ${\bm H}$, respectively.
The quasiparticle scattering rate inside a vortex core $\varGamma$ is given by \cite{higashi2011,nagai2010}
%%%%%%%%%%%%%%%%%%%%%%%%%%%%%
\begin{eqnarray}
\varGamma(\varepsilon)
&=&
\frac{\pi}{2} \varGamma_{\rm n}
\Bigg\langle
\Bigg\langle
\bigl(1-\mathop{\mathrm{sgn}}\nolimits[d({\bm k}_{\rm F})d({\bm k}^\prime_{\rm F})] \cos\Theta  \bigr) 
\nonumber \\
& &\qquad \times
\frac{1}{\vert\sin\Theta\vert}
\frac
{\vert {\bm v}_{\rm F \perp}({\bm k}^\prime_{\rm F}) \vert}
{\vert \bm{v}_{\rm F \perp}({\bm k}_{\rm F}) \vert}
\frac
{\vert d({\bm k}_{\rm F}) \vert}
{\vert d({\bm k}^\prime_{\rm F})\vert}
\nonumber \\
& &\qquad \times
e^{-u(s_0,{\bm k}_{\rm F})}
e^{-u(s^\prime_0,{\bm k}^\prime_{\rm F})}
\Bigg\rangle_{\rm FS^\prime} \Bigg\rangle_{\rm FS},
\label{qp scattering rate}
\end{eqnarray}
%%%%%%%%%%%%%%%%%%%%%%%%%%%%%%
where $\varGamma_{\rm n}$ is the scattering rate in the normal state,
$
\Theta({\bm k}_{\rm F}, {\bm k}^\prime_{\rm F})
\equiv
\theta_v({\bm k}_{\rm F})-\theta_{v^\prime}({\bm k}^\prime_{\rm F})
$
is the scattering angle and
$
u(s,{\bm k}_{\rm F})
=
(2\vert d({\bm k}_{\rm F}) \vert/
\vert {\bm v}_{\rm F \perp}({\bm k}_{\rm F}) \vert)
\int_0^{\vert s \vert}ds^\prime
{\Delta}(s^\prime)
$
with
$\Delta(s^\prime)=\Delta_\infty \tanh(s^\prime/\xi)$.
$s^\prime$ is the real space coordinate along the QP trajectory.
One can obtain further information on the expression of $\varGamma$ in Ref.~\cite{higashi2011,nagai2010}.

We consider two model FSs I, II with in-plane anisotropy.
The model FS I is characterized by the energy dispersion
$
\epsilon_{\rm I}({\bm k})
=
-\mu_{\rm I}-2t\left\{ \cos(k_x a) +\cos(k_y a) \right\}
+k^2_z/(2m),
$
where $t$ and $a$ are the hopping integral and the lattice constant, respectively.
$\mu_{\rm I}$ is the chemical potential.
The dispersion in the $k_x - k_y$ plane is given by the the tight-binding (TB) model
and that in the $k_z$ direction is free electron model.
As characteristics of the TB model,
there are Van Hove singularities in the direction of $(\pi,0)$ and $(0,\pi)$,
at which $|{\bm \nabla}_{\bm k} \epsilon({\bm{k}})|=0$ \cite{sadowski}.
In addition to this point,
the anisotropy of the FS grows larger gradually with increasing the chemical potential
below the half filling.
The model FS II is given by the anisotropic dispersion
$
\epsilon_{\rm II}({\bm k})
=
-\mu_{\rm II}+1/(2m)
\left\{
k^2_x+k^2_y
+a^2(k^4_x+k^4_y)/2+3a^2k^2_x k^2_y
+k^2_z
\right\}
$
\cite{vekhter2008,comment}.
$m$ is the mass of charge.
In numerical calculation,
we set the parameter $mt a^2=1$.

For an isotropic FS,
the position on the FS is identified by the azimuthal and the polar angle $(\phi_k,\theta_k)$.
However, in anisotropic FSs,
it is identified by $\phi_k$, $\theta_k$ and the Fermi radius $|{\bm k}_{\rm F}(\phi_k,\theta_k)|$.
We can parametrize the Fermi wave numbers in spherical coordinates: 
$
k_{{\rm F} x} = |{\bm k}_{\rm F}(\phi_k,\theta_k)| \cos \phi_k \sin \theta_k,
k_{{\rm F} y} = |{\bm k}_{\rm F}(\phi_k,\theta_k)| \sin \phi_k \sin \theta_k,
k_{{\rm F} z} = |{\bm k}_{\rm F}(\phi_k,\theta_k)| \cos \theta_k.
$
Substituting these Fermi wave numbers into the above two dispersions $\epsilon_{\rm I}({\bm k})$ and $\epsilon_{\rm II}({\bm k})$,
and using a bisection method,
we can determine numerically $|{\bm k}_{\rm F}(\phi_k,\theta_k)|$
such that $\epsilon({\bm k})=0$.

For FS I, the absolute value of the  Fermi velocity is
%%%%%%%%%%%%%%%%%%%%%%%%%%%%%%%%%%%%%%%%%%%%%%%%%%%%%%%
\begin{equation}
|{\bm v}_{\rm F}(\phi_k,\theta_k)| = 2ta \sqrt{\sin^2(k_{{\rm F}x}a)+\sin^2(k_{{\rm F}y}a)+\frac{1}{4(mt a^2)^2}k^2_{{\rm F}z}}
\end{equation}
%%%%%%%%%%%%%%%%%%%%%%%%%%%%%%%%%%%%%%%%%%%%%%%%%%%%%%%
and for FS II,
%%%%%%%%%%%%%%%%%%%%%%%%%%%%%%%%%%%%%%%%%%%%%%%%%%%%%%%
\begin{eqnarray}
|{\bm v}_{\rm F}(\phi_k,\theta_k)| &=& \frac{1}{m}
\Bigl\{
 k^2_{{\rm F}x}\left(1+a^2 k^2_{{\rm F}x}+3a^2 k^2_{{\rm F}y}\right)^2
\nonumber \\
& &
+k^2_{{\rm F}y}\left(1+a^2k^2_{{\rm F}y}+3a^2k^2_{{\rm F}x}\right)^2+k^2_{{\rm F}z}
\Bigr\}^{1/2}.
\end{eqnarray}
%
%\begin{equation}
%|{\bm v}_{\rm F}(\phi_k,\theta_k)| = \frac{1}{m}
%\sqrt{k^2_{{\rm F}x}\left(1+a^2 k^2_{{\rm F}x}+3a^2 k^2_{{\rm F}y}\right)^2
%+k^2_{{\rm F}y}\left(1+a^2k^2_{{\rm F}y}+3a^2k^2_{{\rm F}x}\right)^2+k^2_{{\rm F}z}}.
%\end{equation}
%%%%%%%%%%%%%%%%%%%%%%%%%%%%%%%%%%%%%%%%%%%%%%%%%%%%%%%
When integrating Eq.~(\ref{qp scattering rate}) numerically,
we need a relation beween ${\bm v}_{\rm F \perp}(\phi_k,\theta_k)$ and ${\bm v}_{\rm F}(\phi_k,\theta_k)$ \cite{higashi2011}.
The component of ${\bm v}_{\rm F}(\phi_k,\theta_k)$ projected onto the plane perpendicular to ${\bm H}$ is given by
%%%%%%%%%%%%%%%%%%%%%%%%%%%%%%%%%%%%%%%%%%%%%%%%%%%%%%%
\begin{eqnarray}
|{\bm v}_{\rm F \perp}(\phi_k,\theta_k)| &=& |{\bm v}_{\rm F}(\phi_k,\theta_k)| 
\nonumber \\
& & \times \sqrt{\cos^2 \theta_k + \sin^2 \theta_k \sin^2(\phi_k-\alpha_{\rm M})},\\
\cos \theta_v ({\bm k}_{\rm F}) &=& -\frac{|{\bm v}_{\rm F}(\phi_k,\theta_k)|}{|{\bm v}_{\rm F \perp}(\phi_k,\theta_k)|} \cos \theta_k,\\
\sin \theta_v ({\bm k}_{\rm F}) &=& \frac{|{\bm v}_{\rm F}(\phi_k,\theta_k)|}{|{\bm v}_{\rm F \perp}(\phi_k,\theta_k)|} \sin \theta_k \sin(\phi_k-\alpha_{\rm M}),
\end{eqnarray}
%%%%%%%%%%%%%%%%%%%%%%%%%%%%%%%%%%%%%%%%%%%%%%%%%%%%%%%%
where $\theta_v ({\bm k}_{\rm F})$ is the angle of the QP trajectory measured from the ${\bm a}_{\rm M}$-axis.
The ${\bm a}_{\rm M}$ - ${\bm b}_{\rm M}$ plane is perpendicular to ${\bm H}$ (${\bm c}_{\rm M}~||~{\bm H}$).
$\alpha_{\rm M}$ is the magnetic field angle measured from the $(\pi,0)$ direction.

In order to investigate the relation between the anisotropy of the pair potential and that of the FSs,
we consider the following four model pair potentials:
(i) Line-node $s_{|x^2-y^2|}$-wave: $d({\bm k}_{\rm F})=|\cos(2\phi_k) \sin^2 \theta_k|$,
(ii) Line-node $s_{|xy|}$-wave: $d({\bm k}_{\rm F})=|\sin(2\phi_k) \sin^2 \theta_k|$,
(iii) $d_{x^2-y^2}$-wave: $d({\bm k}_{\rm F})=\cos(2\phi_k) \sin^2 \theta_k$,
and (iv) $d_{xy}$-wave: $d({\bm k}_{\rm F})=\sin(2\phi_k) \sin^2 \theta_k$.
While $s_{|x^2-y^2|}$ ($s_{|xy|}$)-wave and $d_{x^2-y^2}$ ($d_{xy}$)-wave pair potentials
have the same anisotropy,
only $d_{x^2-y^2}$ ($d_{xy}$)-wave one has the sign change.
$s_{|x^2-y^2|}$ ($d_{x^2-y^2}$)-wave pair potential coincides with $s_{|xy|}$ ($d_{xy}$)-wave one
when the pair potential is rotated by $\pi/4$ [rad].

% #########################################################################
% #########################################################################
\section{Results}
%%%%%%%%%%%%%%%%%%%%%%%%%%%%%%%%%
\begin{figure}[t]
\begin{center}
\begin{tabular}{p{70mm}p{70mm}}
      \resizebox{70mm}{!}{\includegraphics{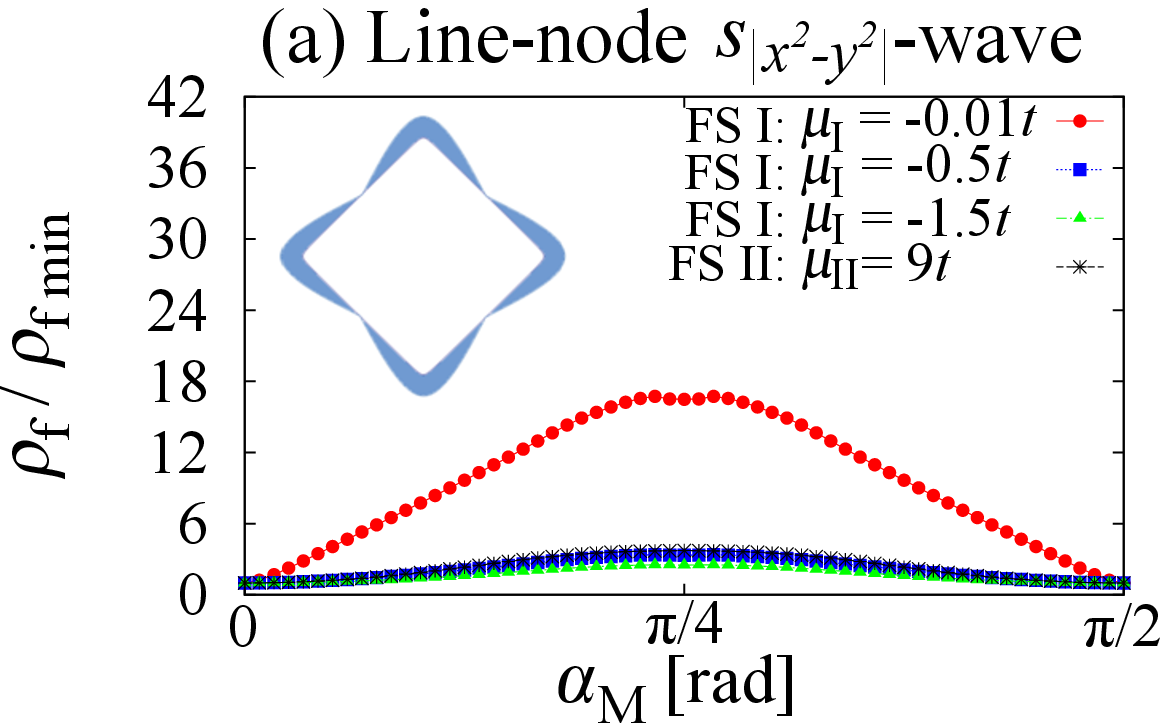}} \\
      \resizebox{70mm}{!}{\includegraphics{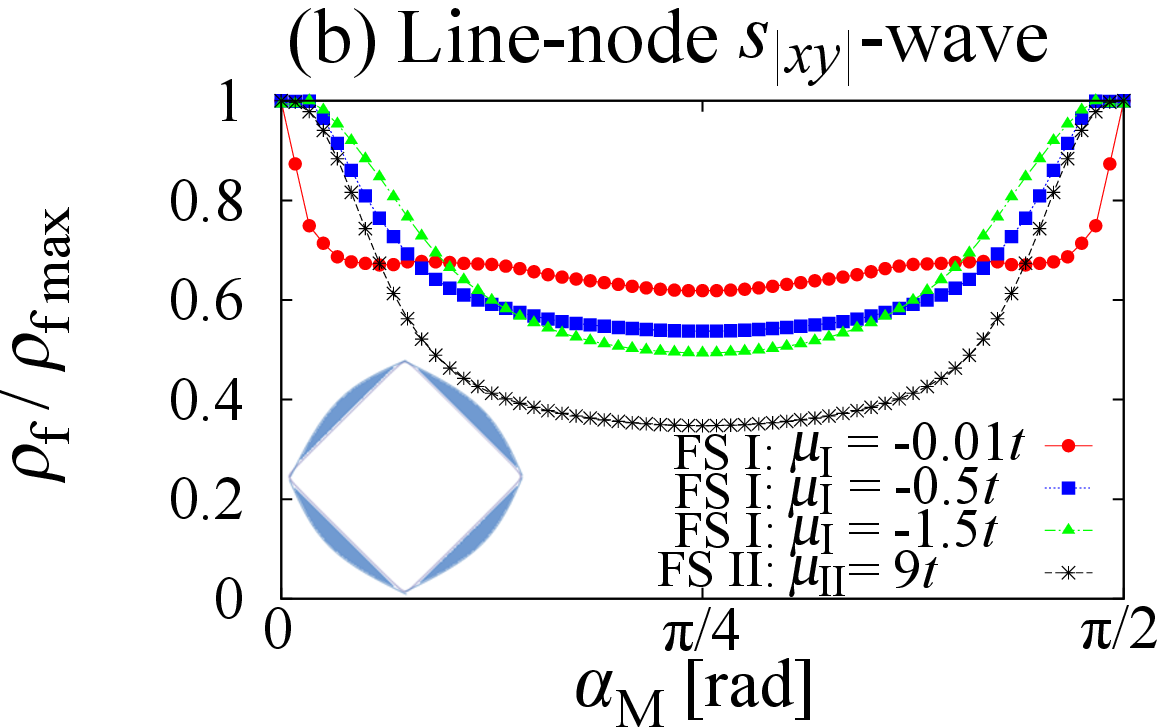}} 
    \end{tabular}
\caption{\label{Fig.1}
The field angle $\alpha_{\rm M}$ dependence of
the flux-flow resistivity $\rho_{\rm f}$
in the case of the ($a$) line-node $s_{|x^2-y^2|}$-wave pair
and the ($b$) line-node $s_{|xy|}$-wave one.
The temperature is set to $T=0.35T_{\rm c}$.
Each curve is plotted for the different chemical potential.
For FS II, the chemical potential is set to $\mu_{\rm II}=9t$.
The vertical axis is normalized by ($a$) minimum value
and ($b$) maximum value for each curve.
}
\end{center}
\end{figure}
%%%%%%%%%%%%%%%%%%%%%%%%%%%%%%%%%
%%%%%%%%%%%%%%%%%%%%%%%%%%%%%%%%%
\begin{figure}[t]
\begin{center}
    \begin{tabular}{p{70mm}p{70mm}}
      \resizebox{70mm}{!}{\includegraphics{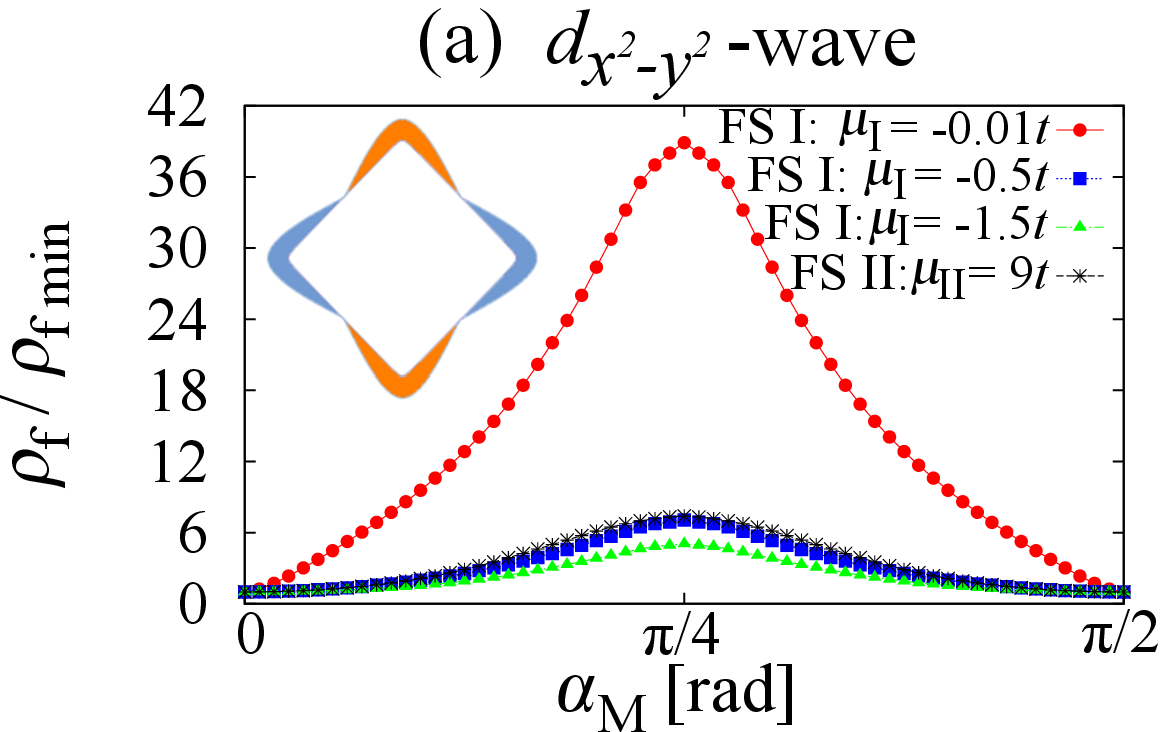}} \\
      \resizebox{70mm}{!}{\includegraphics{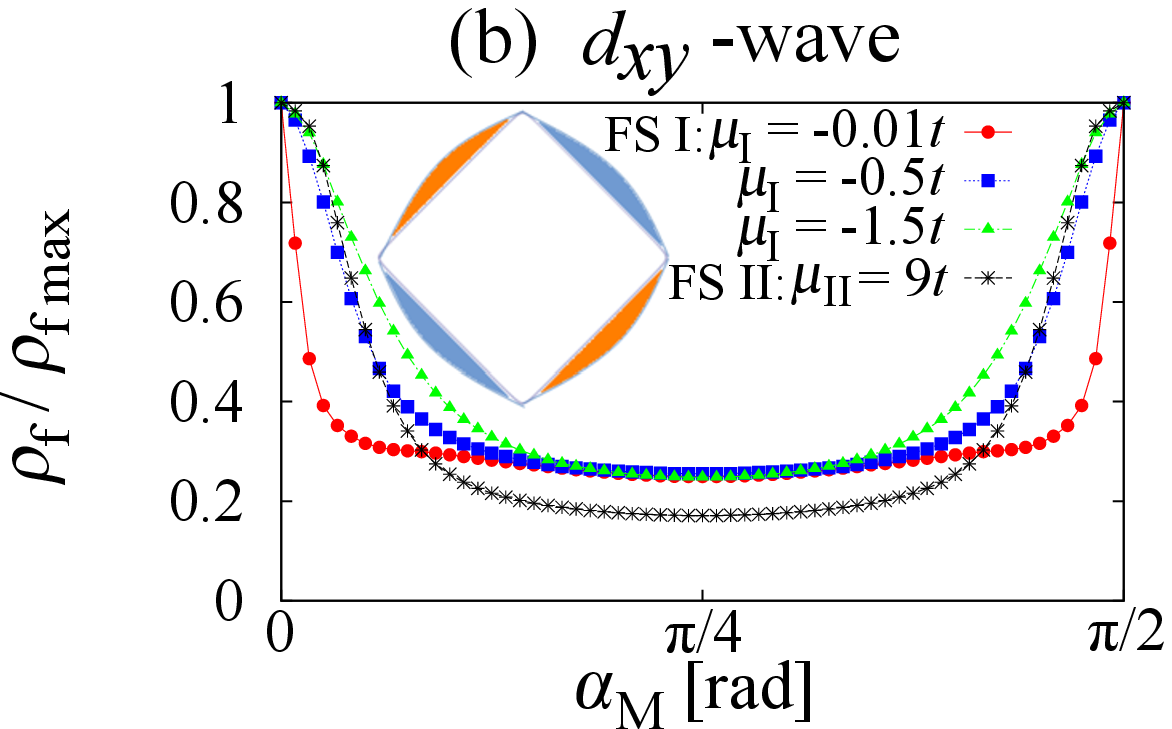}} 
    \end{tabular}
\caption{\label{Fig.2}
The field angle $\alpha_{\rm M}$ dependence of
the flux-flow resistivity $\rho_{\rm f}$
in the case of the ($a$) $d_{|x^2-y^2|}$-wave pair
and the ($b$) $d_{|xy|}$-wave one.
The temperature is set to $T=0.35T_{\rm c}$.
Each curve is plotted for the different chemical potential.
For FS II, the chemical potential is set to $\mu_{\rm II}=9t$.
The vertical axis is normalized by ($a$) minimum value
and ($b$) maximum value for each curve.
}
\end{center}
\end{figure}
%%%%%%%%%%%%%%%%%%%%%%%%%%%%%%%%%
%%%%%%%%%%%%%%%%%%%%%%%%%%%%%%%%%
In Figs.~\ref{Fig.1} and \ref{Fig.2},
we show the numerical results of the field-angle $\alpha_{\rm M}$ dependece of the flux-flow resistivity $\rho_{\rm f}$
for the model FS I and II.
The inset shows the schematic figure of the pair potential on the anisotropic FS.
We fix the temperature in this calculation at $T=0.35T_{\rm c}$.
Each plot corresponds to different chemical potential.
For the energy dispersion II,
the chemical potential $\mu_{\rm II}$ is fixed to $\mu_{\rm II}=9t$.

In Figs.~\ref{Fig.1}(a) and \ref{Fig.1}(b),
we can see that $\rho_{\rm f}$ has the maximum value
when ${\bm H}$ is applied parallel to the gap-node direction.
This behavior in the $s_{|x^2-y^2|}$-wave pair is consistent with the result
for both an isotropic and an uniaxially anisotropic FS \cite{higashi}.
The same behavior is seen also for the $d_{x^2-y^2}$ and the $d_{xy}$-wave pair [see Fig.~\ref{Fig.2}(a) and \ref{Fig.2}(b)].
However,
the oscillation amplitude of $\rho_{\rm f}(\alpha_{\rm M})$ for the $d_{x^2-y^2}$ and the $d_{xy}$-wave pair
becomes larger than that for the $s_{|x^2-y^2|}$ and the $s_{|xy|}$-wave one.
In addition,
for the $d_{x^2-y^2}$-wave pair,
a rather sharper peak appears when ${\bm H}~||~\pi/4$.
These characteristics originate from the sign-change in the pair potential because its amplitude is the same
between the $s_{|x^2-y^2|}$ and the $d_{x^2-y^2}$ (or the $s_{|xy|}$ and the $d_{xy}$)-wave pair.

We notice that the curve of $\rho_{\rm f}(\alpha_{\rm M})$ for FS I
approaches the curve for FS II
with decreasing the FS anisotropy ($\mu_{\rm I}=-0.01t\rightarrow -1.5t$) in Figs.~\ref{Fig.1} and \ref{Fig.2}.
The anisotropy of FS II with $\mu_{\rm II}=9t$ is almost the same as that of FS I with $\mu_{\rm I}=-1.5t$.
The cusp-like sharp peak appearing in an isotropic FS for the $d_{x^2-y^2}$-wave pair \cite{higashi} is not observed
in the case of these in-plane anisotropic FSs.
For any pairing states,
when ${\bm H}~||$ gap node,
the peak becomes sharp
with increasing the anisotropy of the FS.
We consider that this behavior comes from the anisotropy of the FS.
The important point is that
one can detect the gap-node direction
from the field-angle dependence of the flux-flow resistivity
even if there is an anisotropy of a FS.
That is,
a peak of $\rho_{\rm f}(\alpha_{\rm M})$ appears in the gap-node direction
irrespective of a FS anisotropy.

% #########################################################################
% #########################################################################
\section{Summary}
We investigated the field-angle dependence of the flux-flow resistivity
for two models of FSs.
As a result,
we find that the maximum value of $\rho_{\rm f}(\alpha_{\rm M})$
always appears when ${\bm H}$ is oriented parallel to the gap-node direction
even if the FS has an anisotropy.
This result is irrespective of the relation between the anisotopy of the pair potential and that of the FS
[compare Fig.~\ref{Fig.1}(a) with \ref{Fig.1}(b) and Fig.~\ref{Fig.2}(a) with \ref{Fig.2}(b)].

% #########################################################################
% #########################################################################
\section*{Acknowledgments}
The authors thank N. Nakai, H. Suematsu, T. Okada, S. Yasuzuka, Y. Kato, K. Izawa, M. Kato, and A. Maeda for helpful discussions.



%% References

%\section*{References}
\bibliographystyle{elsarticle-num}
\bibliography{<your-bib-database>}




%%%%%%%%%%%%%%%%%%%%%%%%%%%%%%%%%%%%%%%%%%%%%%%%%%%%%%%%%%%%%%%%%%%%%%%%%%%%%%%%%%%%%%%%%%%%%%%%%%%%%%%%%%%%%%%%%%%%%%%%%%%%%%%%%%%%%%%%
%%%%%%%%%%%%%%%%%%%%%%%%%%%%%%%%%%%%%%%%%%%%%%%%%%%%%%%%%%%%%%%%%%%%%%%%%%%%%%%%%%%%%%%%%%%%%%%%%%%%%%%%%%%%%%%%%%%%%%%%%%%%%%%%%%%%%%%%



% \end{thebibliography}

\end{document}